\RequirePackage[2020-02-02]{latexrelease}

\documentclass[%
 reprint,
prstab,
]{revtex4-2}

\usepackage{graphicx}
\usepackage{dcolumn}
\usepackage{bm}
\usepackage{import}
\usepackage[separate-uncertainty=true]{siunitx}
\usepackage{amsmath}
\usepackage{xcolor}
\usepackage{ulem}
\usepackage[caption=false]{subfig}

\begin{document}

\preprint{APS/123-QED}

\title{Measurement of the decay of laser-driven linear plasma wakefields}
\author{J. Jonnerby$^{1,\dagger}$, A. von Boetticher$^1$, J. Holloway, L. Corner$^2$,  A. Picksley$^{1,\ddagger}$, A. J. Ross$^1$, R. J. Shalloo$^{3, \mathsection}$, C. Thornton$^4$, N. Bourgeois$^4$, R. Walczak$^{1, \P}$, S. M. Hooker$^1$}

 \email{simon.hooker@physics.ox.ac.uk}
\affiliation{$^1$John Adams Institute for Accelerator Science and Department of Physics,University of Oxford, Denys Wilkinson Building, Keble Road, Oxford OX1 3RH, United Kingdom.\\
$^2$Cockcroft Institute of Accelerator Science, University of Liverpool, Liverpool, United Kingdom\\
$^3$John Adams Institute for Accelerator Science, Imperial College London, London, United Kingdom\\
$^4$Central Laser Facility, STFC Rutherford Appleton Laboratory, Didcot, United Kingdom\\
$^\dagger$Now at NIHR Health Protection Research Unit in Respiratory Infections, Imperial College London, London, United Kingdom\\
$^\ddagger$Now at BELLA Center, Lawrence Berkeley National Lab, Berkeley, California, United States\\
$^\mathsection$Now at Deutsches Elektronen-Synchrotron DESY: Hamburg, Germany\\
$^\P$Somerville College, Woodstock Road, Oxford OX2 6HD, United Kingdom
}%

\date{\today}

\begin{abstract}
We present the first measurements of the temporal decay rate of one-dimensional, linear Langmuir waves excited by an ultra-short laser pulse. Langmuir waves with relative amplitudes of approximately $6\%$ were driven by \SI{1.7}{J}, \SI{50}{fs} laser pulses in hydrogen and deuterium plasmas of density $n_{e0} = \SI{8.4E17}{cm^{-3}}$. The wakefield lifetimes were measured to be $\tau^\mathrm{H_2}_\mathrm{wf} = (9\pm2)$ ps and $\tau^\mathrm{D_2}_\mathrm{wf} = (16\pm8)$ ps respectively for hydrogen and deuterium. The experimental results were found to be in good agreement with 2D particle-in-cell simulations. In addition to being of fundamental interest, these results are particularly relevant to the development of laser wakefield accelerators (LWFAs) and wakefield acceleration schemes using multiple pulses, such as multi-pulse laser wakefield accelerators (MP-LWFAs).
\end{abstract}

\maketitle
\section{Introduction}\label{sec:intro}

Next-generation plasma wakefield particle accelerators use charged beams or laser pulses to excite Langmuir waves that can support accelerating gradients on the order of GeV/cm \cite{Tajima_1979}. Considerable progress has been made in this sphere in recent years, including, for example: the acceleration to multi-GeV-scale energies in wakefields driven by laser pulses \cite{Leemans_2006, Gonsalves2019}, electron bunches \cite{Yakimenko2019, pompili_2021}, and by long proton bunches \cite{Adli2018}; and applications of plasma-accelerated beams to the generation of radiation \cite{Albert2018}, and the first demonstrations of gain in free-electron-laser experiments \cite{Wang2021, Ferrario2022}.

In the original concept \cite{Tajima_1979} of a laser-driven plasma accelerator, the driving laser pulse had a duration shorter than the plasma period $T_{pe}=2\pi/\omega_{pe}$, where $\omega_{pe} =(n_{e0} e^2/m_e \epsilon_0)^{1/2}$, and in which $e$ is the electron charge, $m_e$ is the electron mass, and $\epsilon_0$ is the vacuum permittivity.  Many important results have been obtained in this regime, for both laser- and particle-beam-driven plasma accelerators, and this regime continues to be a major focus of research worldwide. However, it is also possible to drive the plasma wakefield with: (i) a train of short drive pulses, spaced by $T_{pe}$;  or (ii) by a drive pulse that is long compared to $T_{pe}$, but with a temporal intensity profile that is modulated with a period of $T_{pe}$ \cite{Hooker_2014, Cowley2017a, Yakimenko2019, Adli2018, Muggli2010}.

We recently extended this latter concept by proposing a new method \cite{Jakobbson:2021} for generating the required pulse train: frequency modulation of a long laser pulse by a plasma wave driven by a short, low-energy seed pulse, followed by temporal compression in a dispersive optical system. Simulations of this scheme show that electrons could be accelerated to \SI{0.65}{GeV} in a plasma stage driven by a pulse train generated by a \SI{1.7}{J}, \SI{1}{ps} drive pulse of the type which could be provided by a kilohertz repetition rate thin-disk laser \cite{Tunnermann2010}. For plasma accelerators driven by long ($\tau_{\text{drive}} \gg T_{pe}$) drivers, it is important to understand the extent to which the amplitude of the plasma wave decays over the total duration of the driver.

Theoretical studies of plasma dynamics have shown that interactions between the oscillating electrons and the background ions can lead to the growth of instabilities which dissipate the Langmuir wave energy into higher-order daughter waves \cite{Mora1988}. Ultimately, these instabilities lead to the decay of the wakefields and heating of the plasma. In this paper, we present the results of an experimental investigation of the wakefield decay rate in a parameter regime that is relevant for several current and future plasma acceleration schemes, such as plasma wakefield acceleration (PWFA) \cite{Adli2018}, laser wakefield acceleration (LWFA) \cite{Leemans_2006} and multi-pulse laser wakefield acceleration (MP-LWFA) \cite{Hooker_2014}. We compare our measured results with particle-in-cell simulations, and show that there is good agreement between theory, simulations, and measurements. 

\begin{table*}[t]
\begin{center}
\caption{Comparison of key parameters for several experiments to measure the decay of laser-driven plasma waves.}\label{tab:expt_parameters}

\begin{tabular}{lrcrrrrrrrccc}\toprule
Regime &Target &Plasma &$\tau_L$ &$n_{e0}$ / $\SI{}{\per\cubic\cm}$ &$\delta n_e/n_{e0}$ &$T$ / eV &$W$ &$\tau_L/T_{pe}$ &$\left(\lambda_{p}/\pi\sigma\right)^2$ &$\tau_{wf}/T_{pe}$ &$\tau_{wf}$ [ps] &Ref \\
\hline
LBWA &Cell &D$_2$ &$\SI{160}{\pico\s}$ &$\SI{1.07e17}{}$ &$0.1$ &$20$ &$270$ &$623$ &$0.11$ &$61\pm59$ &$20.6\pm 20$ &\cite{Moulin1994} \\
SM-LWFA &Jet &He &$$\SI{400}{\femto\s}$$ &$\SI{3e19}{}$ &$0.15$ &$2500$ &$4.6$ &$20$ &$0.023$ &$132\pm14$ &$1.9\pm0.2$ &\cite{LeBlanc1996} \\
SM-LWFA &Jet &He &$\SI{400}{\femto\s}$ &$\SI{3.7e19}{}$ &$0.1$ &$1000$ &$5.11$ &$21$ &$0.033$ &$139$ &$1.8$ &\cite{Chen2000} \\
SM-LWFA &Jet &H$_2$ &$\SI{400}{\femto\s}$ &$\SI{1e19}{}$ &$0.1$ &$10$ &$255$ &$11$ &$0.62$ &$142\pm28$ &$6\pm1$ &\cite{Ting1996} \\
SM-LWFA &Jet &He &$\SI{400}{\femto\s}$ &$\SI{1e19}{}$ &$0.1$ &$10$ &$255$ &$11$ &$0.62$ &$142\pm28$ &$6\pm1$ &\cite{Ting1996} \\
& & & & & & & & & & & & \\
LWFA &Cell &He &$\SI{120}{\femto\s}$ &$\SI{1e17}{}$ &$0.1$ &$14$ &$36$ &$0.34$ &$31.4$ &$33\substack{+33\\-8}$ &$8.3\substack{+8\\-2}$ &\cite{Marques1997a} \\
LWFA &Jet &He &$\SI{52}{\femto\s}$ &$\SI{7.4e17}{}$ &$0.75$ &$13$ &$1150$ &$0.4$ &$1.92$ &$9.7$ &$1.3$ &\cite{Kotaki2002} \\
& & & & & & & & & & & & \\
LWFA &Cell &H$_2$ &$\SI{48.9\pm6.3}{\femto\s}$ &$\SI{8.4e17}{}$ &$0.06$ &$2$ &$705$ &$0.4$ &$0.042$ &$84\pm25$ &$9\pm2$ &This work \\
LWFA &Cell &D$_2$ &$\SI{48.9\pm6.3}{\femto\s}$ &$\SI{8.4e17}{}$ &$0.04$ &$2$ &$262$ &$0.4$ &$0.042$ &$134\pm63$ &$16\pm8$ &This work\\
\toprule
\end{tabular}
\end{center}
\end{table*}
We first establish the key laser and plasma parameters which determine the regime in which a laser-plasma accelerator operates. When the quiver velocity of the plasma electrons in the field of the driving laser is non-relativistic, the wakefield is approximately sinusoidal, and is said to be in the linear regime. For a single, short driving laser pulse, this corresponds to a peak normalised vector potential $a_0 < 1$, where $a_0 = eE/m_ec\omega_0$, $E$ the laser electric field strength, and $\omega_0$ the laser frequency.  In this regime, and for the case of a driving laser pulse with Gaussian temporal and transverse intensity profiles, the relative amplitude of the plasma wave $\delta n_e/n_{e0}$ is given by the sum of the of the relative amplitudes of the radial and longitudinal wakefields  \cite{Marques1997a},
\begin{align}
  \frac{\delta n_e}{n_{e0}} =& \frac{\delta n_r}{n_{e0}} + \frac{\delta n_z}{n_{e0}} \nonumber\\
  =& A\left[\underbrace{1}_{\text{Long.}}+\underbrace{\left(\frac{2 c}{\omega_{pe} \sigma}\right)^{2}\left(1-\frac{r^{2}}{\sigma^{2}}\right)}_{\text{Radial}}\right]\times\nonumber\\
  &\exp \left(\frac{-r^{2}}{\sigma^{2}}\right)\sin \left(\omega_{pe}\zeta\right)\\
  A=&\frac{I \sqrt{\pi}}{c^{3} n_{c} m_{e}}\left(\frac{\omega_{pe} \tau_{L}}{2}\right) \exp \left[-\left(\frac{\omega_{pe} \tau_{L}}{2}\right)^{2}\right],
  \label{eq:wake_amp_dn}
\end{align}
where $r$ is the radial distance from the propagation axis of the drive laser, $\zeta$ is the temporal delay after the peak of the drive pulse, $\omega_{pe}=(n_{e0}e^2/m_e\epsilon_0)^{1/2}$ is the plasma frequency, $n_{e0}$ is the plasma electron density, $n_c=\epsilon_0m_e\omega^2/e^2$ is the critical density, and $\omega$, $\sigma$, $\tau_L$ and $I$ are respectively the angular frequency, beam radius at the $1/e^2$ intensity, the duration (defined as the half width at $1/e^2$ intensity) and peak intensity of the driving laser pulse. The ratio of the radial to the longitudinal wakefield components at $r=0$ is $\delta n_r/\delta n_z|_{r=0} =  (2 c /\omega_{pe} \sigma)^2 = (\lambda_p / \pi \sigma)^2$, where $\delta n_r/\delta n_z\gg1$ indicates a predominantly radial wakefield and $\delta n_r/\delta n_z\ll1$ indicates a longitudinal wakefield, which we will refer to as a one-dimensional wakefield.

The mechanisms responsible for the decay of laser-driven plasma waves have been investigated in several earlier studies. Marqu\`es et al. \cite{Marques1997a} found that radial-dominated and longitudinal-dominated wakefields can have different decay mechanisms. Longitudinal wakefields decay through collisions \cite{Banks2016}, Landau damping \cite{bellan_2006}, beam loading by accelerated particles \cite{LeBlanc1996}, and the modulational instability \cite{Mora1988}. The  growth rate of the modulational instability is expected to be much greater than the collisional or Landau damping mechanisms, and hence will usually dominate the decay in the case when beam-loading is not significant. If the total charge trapped and accelerated by the wakefield is large, then beam loading becomes important, and can be the leading cause of the wakefield decay \cite{LeBlanc1996}. Radial wakefields can decay via an additional mechanism. When the radial plasma density is non-uniform, electrons at different radial trajectories have different oscillation periods, which leads to a loss of coherence of the plasma oscillation. This can happen e.g. via a final radial extent of the drive laser or beam, or in pre-formed plasma channels \cite{shalloo_low-density_2019, picksley_guiding_2020, picksley_meter-scale_2020, alejo_demonstration_2022}.

In addition to the ratio $\delta n_r / \delta n_z$, two other parameters are important in determining the mechanisms responsible for, and the rate of, wakefield decay. First, the ratio of the energy density of the Langmuir wave to the thermal energy density, $W = (v_L/v_t)^2 =  \epsilon_0 |\mathbf{E_L}|^2/2n_{e0} k_B T_e$, where $v_L = e E_L/m_e\omega_{pe}$, $v_t = (k_BT_e/m_e)^{1/2}$, $E_L$ is the electric field strength of the wakefield, $k_B$ is the Boltzmann constant, and $T_e$ is the electron plasma temperature \cite{Mora1988}. The parameter $W$ determines the growth rate of the modulational instability and delineates the strong-field regime ($W\gg 1$) from the weak-field regime ($W\lesssim 1$). Second, the ratio of the drive pulse length to the plasma ion period $\tau_L/T_{pi}$, where $T_{pi} = 2 \pi / \omega_{pi}$, where $\omega_{pi} =(Z n_{e0} e^2/M \epsilon_0)^{1/2}$, and in which $Z$ and $M$ are the charge and mass of the ions respectively. For $\tau_L\gg T_{pi}$ the plasma instabilities driven by ion dynamics co-evolve with the drive laser, and for $\tau_L\ll T_{pi}$ they develop only after the wakefield is excited.

Table I summarises the results of previous measurements of the decay time of laser-driven wakefields. The penultimate and antepenultimate columns of the table give the wakefield decay time $\tau_{wf}$ and the ratio of this to the electron plasma period $T_{pe}$. It should be noted that the precise definition of $\tau_{wf}$ varies between the experiments. In Refs. \cite{Ting1996,Moulin1994} it refers to the total length of the detectable wakefield signal, whereas in Refs. \cite{Marques1997a, Kotaki2002, LeBlanc1996, Chen2000}, and in the present work, it refers to the time taken for the wakefield amplitude to decay to $1/e$ of the maximum amplitude. We also note that the time given in Ref \cite{Moulin1994} refers to the saturation time of the beatwave-driven wakefield, and, for simplicity, we have taken this to be equal to the decay time. It is striking that with the exception of Ref. \cite{Kotaki2002}, the ratio of the wakefield lifetime to the electron plasma period $\tau_{wf}/T_{pe}$ varies by less than a factor of 5 in experiments for which the plasma density and the duration of the drive pulse both vary by more than two orders of magnitude. 

The earlier work summarised in Table I was undertaken in a wide range of laser-plasma accelerator regimes. In the experiment by Moulin et al. \cite{Moulin1994} a wakefield was excited using the `beatwave' (LBWA) scheme in which two long pulses ($\SI{160}{\pico\s}$ and $\SI{90}{\pico\s}$), of angular frequencies $\omega_1$ and $\omega_2$, interfere to generate a beat pattern at the plasma frequency $\omega_{pe} =  \omega_1 - \omega_2$. In that work plasma instabilities therefore developed during the excitation of the wakefield by the drive beam or drivers. The experiments by Leblanc et al. \cite{LeBlanc1996}, Chen et al. \cite{Chen2000}, and Ting et al. \cite{Ting1996} corresponded to the self-modulated laser wakefield (SM-LWFA) regime in which interaction between a long laser pulse ($\tau_L \gg T_{pe}$) and the weak plasma wave it drives causes the laser pulse to become modulated with a period of $T_{pe}$, leading to a nonlinear feedback in which the modulation of the laser pulse and the wakefield amplitude both increase with delay relative to the front of the driving pulse. Hence these experiments correspond to an intermediate regime in which the duration of the drive pulse lies between the electron plasma period and the ion plasma period $T_{pi}$. Finally, Marqu\`es et al. \cite{Marques1997a} and Kotaki et al. \cite{Kotaki2002} performed experiments in the LWFA regime  originally proposed by Tajima and Dawson, in which the wakefield is excited by a laser pulse with $\tau_L < T_{pe}$. These last two experiments both operated in the radial-dominated wakefield regime, and shorter decay times relative to the plasma period were observed compared to the experiments that generated longitudinal-dominated wakefields. With the exception of the work by Kotaki et al. \cite{Kotaki2002}, in all the experiments $\delta{n_e}/n_{e0}\approx 0.1$, i.e. they were all conducted in the linear wakefield regime. It is noteworthy that for the much stronger wakefields ($\delta n_e / n_{e0} \approx 0.75$) studied in \cite{Kotaki2002}, the observed ratio $\tau_{wf} / T_{pe}$ is much smaller than found in experiments operating in the linear regime. 

Before concluding this short review of prior experimental work, we note that wakefield decay and ion motion has also been studied for proton-beam-driven wakefield accelerators  \cite{Vieira_2012}; these results have not been included in Table I owing to the very different  driver and plasma parameters. We also note recent experiments to establish the limits to the repetition rate of PWFAs driven by electron bunches \cite{darcy_recovery_2022}. The topic of the present paper, the time-scale for wakefield decay, is related to the maximum possible repetition rate of a plasma accelerator. However, we emphasize that wakefield decay is just the first step in a complex chain of processes --- that includes wakefield decay, electron-ion recombination, and heat redistribution --- that must be completed before the following drive pulse can be delivered. 

To our knowledge, the new work presented in the present paper is the first measurement of the decay rate of a one-dimensional ($\delta n_r/\delta n_z\ll1$), linear wakefield ($a_0\sim 0.5$) in the short-pulse LWFA regime ($\tau_L / T_{pe} \approx 0.4$). This short-pulse regime is relevant for future plasma wakefield facilities \cite{Assmann2019, ALEGROcollaboration2019, Leemans2010} (although we note that some of these are expected to operate in the quasi-linear regime ($a_0\sim 1$ for a single pulse) e.g.\  \cite{Assmann2019}), and it is also relevant to alternative schemes, such as MP-LWFAs \cite{Hooker_2014, Cowley2017a, Jakobbson:2021}.

\section{Measurement of the wakefield lifetime}\label{sec:FDH_exp}

\begin{figure}[tb] 
   \centering
    \includegraphics[width=\columnwidth]{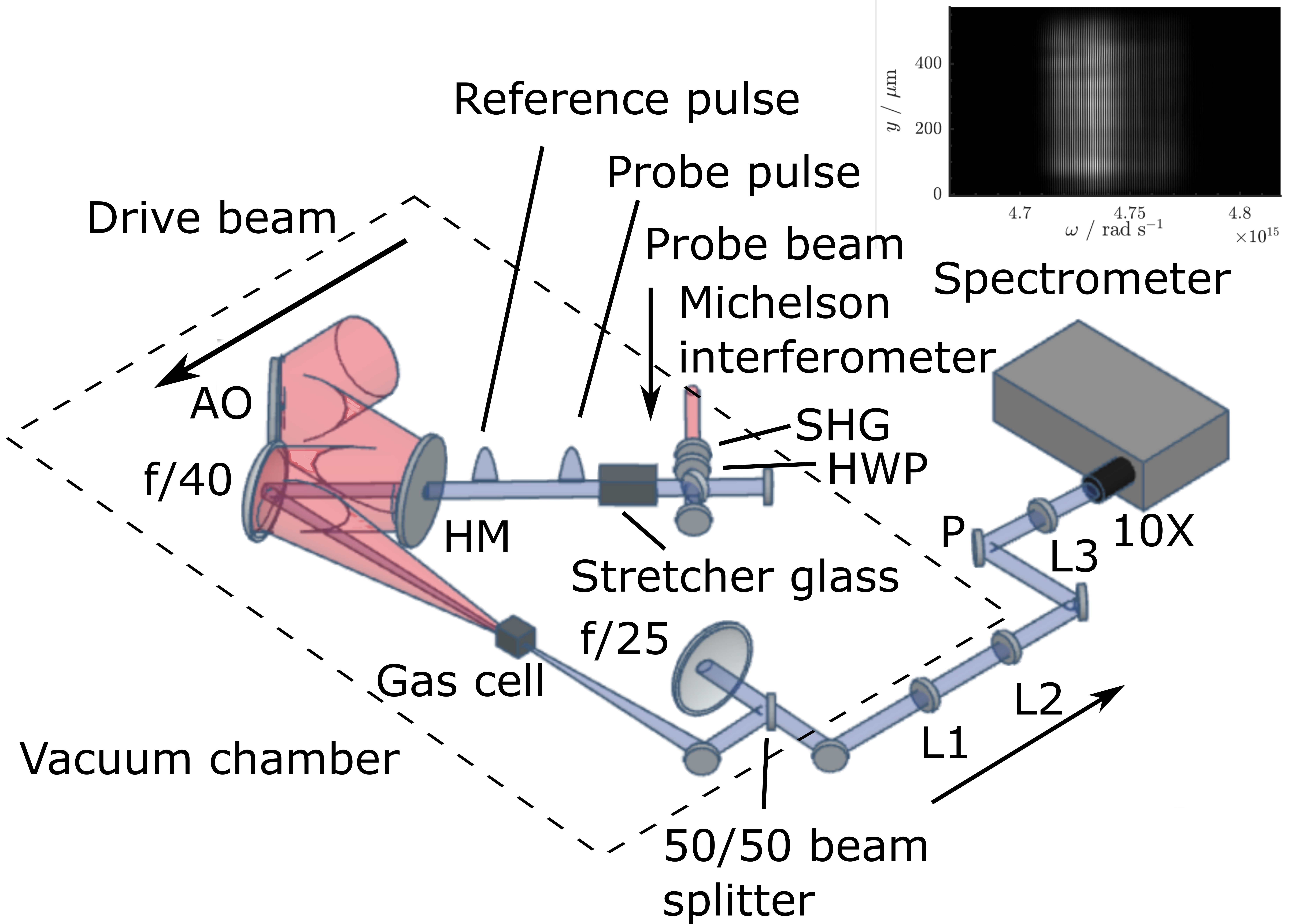}
    \caption{Schematic drawing of the experimental layout inside the target vacuum chamber. Both beams of the Astra-Gemini TA3 laser were used: one as the drive beam, and the other as the diagnostic probe beam. The  \SI{800}{\nano\metre} beams are shown in red and the \SI{400}{\nano\metre} diagnostic beam is shown in blue. After leaving the gas cell, the diagnostic beam was transported to a \SI{400}{\nano\metre} spectrometer located outside the vacuum chamber. The inset shows an example recording of a wakefield in a spectral interferogram, as captured by the spectrometer camera. AO: Adaptive optic, HM: Holed mirror, HWP: Half-wave plate, L1: f=$500$ mm lens, L2: f=$-100$ mm lens, L3: f=$300$ mm lens, P: polariser, $10$X: microscope objective, SHG: second harmonic generating crystal.}
    \label{fig:layout}
\end{figure} 

Experiments to measure the lifetime of a plasma wakefield in the 1D, linear regime were undertaken at the Central Laser Facility of the Rutherford Appleton Laboratory, using the Astra-Gemini TA3 laser. A schematic illustration of the experiment layout is shown in Fig. 1.

A linearly polarised laser pulse with energy $E=\SI{1.68\pm0.06}{\joule}$, centre wavelength \SI{800}{\nano\metre}, and FWHM pulse duration \SI{48.9\pm6.3}{\femto\s} was used to drive the wakefield. This pulse was focused by an on-axis reflecting paraboloid of focal length $f\approx\SI{6.1}{\metre}$, used at $f/40$, to a spot size ($1/e^2$ intensity radius) of $w_0 = \SI{52.3\pm0.8}{\micro\m}$ at the centre of a gas cell. The peak intensity at the laser focus was $I = \SI{6.5e17}{\W\per\cm\squared}$, corresponding to a peak normalised vector potential of $a_0 = 0.54 \pm 0.18$, with approximately a factor of $0.3$ of the beam energy enclosed within the FWHM beam diameter at focus.

Laser radiation could enter and leave the cell via a pair of coaxial, \SI{400}{\micro\m} diameter pinholes located at each end of the \SI{4}{mm} long gas cell. Gas, either hydrogen (H$_2$) or deuterium (D$_2$), was flowed into the cell in a pulse of duration of approximately $500$ ms; the gas flowed into the surrounding vacuum chamber via the pinholes. For these experiments the cell backing pressure was $P_\mathrm{cell} = \SI{17.0\pm1.2}{\milli\bar}$, corresponding to an electron density $\SI{8.4e17}{\per\cubic\cm}$ (assuming $100$\% ionization).

The amplitude of the plasma wave was measured by frequency domain holography (FDH) \cite{Matlis2006}, analysed with the TESS technique \cite{Matlis2016a, Arran2018a}. In this method, two chirped and stretched diagnostic pulses are generated: (i) a probe pulse, which propagates behind the drive pulse, and acquires a temporally-dependent phase shift from the density modulation of the plasma wave; and (ii) a near-identical reference pulse, which propagates ahead of the drive pulse. This pair of diagnostic pulses was generated by passing a frequency-doubled pick-off from the probe beam through a Michelson interferometer with a path difference corresponding to $\Delta \zeta \approx \SI{6}{ps}$. Each of the pair of pulses thereby created was then frequency-chirped and stretched to a duration of \SI{1.35}{ps} by propagating them through a $160$ mm long piece of glass (BK7). The diagnostic pulses were injected coaxially with the drive beam by directing them through a holed turning mirror, and focused into the gas cell by the same optic used to focus the drive beam.  On leaving the gas cell the diagnostic pulses were separated from the transmitted drive pulse by reflection from a dichroic mirror and imaged onto the entrance slit of a Czerny-Turner spectrometer  to yield a spectral interferogram which was recorded by a CCD camera (Andor Newton DU940N-BU).
The wakefield amplitude was calculated from the captured interferograms using the TESS technique, as follows. Each spectral interferogram was Fourier transformed along its spectral axis to give a spatio-temporal profile. The Fourier-transformed data comprises a zero-frequency (``DC'') band; a sideband located at $t=\Delta \zeta$; and three satellites --- two either side of the sideband, and a third located near the DC band. These satellites arise from the phase-shift imposed on the probe beam by the sinusoidal plasma wave. The satellites are offset from the sideband and have temporal locations given by \cite{Matlis2016a, Arran2018a},
\begin{equation}
\tau = \Delta \zeta \pm \varphi^{(2)}\omega_{pe},
\label{eq:peak_sep}
\end{equation}
where $\varphi^{(2)}$ is the group delay dispersion (GDD) of the probe and reference pulses. For plasma waves with large  amplitudes, higher-order satellites can appear, located at $\tau = \Delta \zeta \pm m\varphi^{(2)}\omega_{pe}$, $m=2, 3, 4,...$, but these higher-order satellites were not observed in this experiment.
The relative amplitude of the wakefield at delay $\zeta$ can be found from the ratio $r$ of the satellite amplitude to that of the sideband since this is given by \cite{Matlis2016a, Arran2018a},
\begin{align}
&r = \mathcal{F}(\omega_{pe})\frac{J_1(\phi_0)}{J_0(\phi_0)}
\end{align}
where,
\begin{align}
&\phi_0 = \frac{\omega_{pe}^2L}{2\omega_\mathrm{probe} c}\frac{\delta n_e}{n_{e0}}.\label{eq:phi0}
\end{align}
and where $J_0$ and $J_1$ are Bessel functions of the first kind, $\mathcal{F}(\omega_{pe})$ a spectral overlap function (see Supplemental Material \cite{supp}), $L$ is the interaction length, and $\omega_\mathrm{probe}$ is the frequency of the probe laser. By varying the backing pressure, and measuring $\omega_{pe}$ immediately after the drive pulse, it was found that the pressure in the cell was linearly related to the measured backing pressure through $P_\mathrm{cell} = \alpha( P_\mathrm{gauge}-P_0)$, where $\alpha=0.96$ accounts for the fact that the pressure guage was located prior to the gas cell gas inlet and $P_0=3$ mbar (see Supplemental Material \cite{supp}).

\section{Results}

\begin{figure}[tb]
    \centering
    \includegraphics[width=8cm]{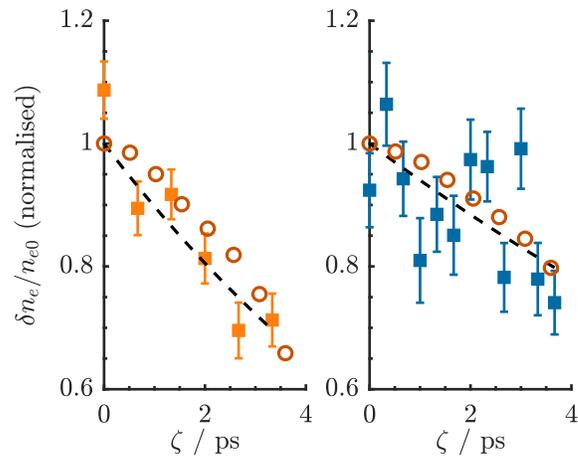}
    \caption{Measured normalised relative wakefield amplitude as a function of delay for: (a) hydrogen; (b) deuterium, recorded with a backing pressure $P_\text{cell} = \SI{17.0\pm1.2}{\milli\bar}$. For each delay are shown the uncertainty-weighted average wakefield amplitude ($\delta n_e(\zeta)/ n_{e0}$) and the standard error. The uncertainty was calculated using the background noise in the Fourier transformed interferograms. The wakefield amplitude calculated from the PIC simulations are shown as open circles. Also shown are fits of the exponential function to the data  as black lines.}
    \label{fig:Decay}
\end{figure}

The wakefield amplitude was measured for a range of delays $\zeta$ in steps of \SI{0.67}{\pico\s}. For each temporal delay, the results of 10 shots were averaged in order to reduce the statistical error in the measured wake amplitude.

Fig. 2 shows, for hydrogen and deuterium, the measured wakefield amplitude as a function of delay, normalised to  $\delta n_e(0)$, where $\delta n_e(0)$ is determined by a fit of the function $\delta n_e(\zeta)  =\delta n_e(0) \exp ( - \zeta / \tau_{wf} )$ to the data. Due to variations in the experimental conditions, the plasma waves driven in deuterium had an initially lower amplitude, which made the error bars relatively larger after normalisation.

Also shown in Fig. 2 are the results of 2D particle-in-cell (PIC) simulations performed with the relativistic particle-in-cell code \texttt{smilei} \citep[Simulation of Matter Irradiated by Light at Extreme Intensities,][]{SMILEI} for the laser and plasma parameters used in the experiment (see Supplemental Material). The wakefield amplitude was calculated from the maximum Fourier amplitude (near $\omega_{pe}$) of the density variation $\delta n (x,y) = \delta n_e (x,y) - \delta n_i (x,y)$ corresponding to the wakefield plasma oscillation, where $\delta n_i (x,y)$ is the ion density. The Fourier amplitude, $A_F$, was converted to the wakefield amplitude, $A_{\text{wake}}$ using the relation $A_{\text{wake}} = 2A_F \Delta x/[\int W_T^{\beta}(x)\ dx]$. Here $W_T^\beta(x)$ is a Tukey window with window parameter $\beta$ (set to $\beta$=0.2), and $\Delta x = 26.7 \ \mathrm{nm}$, the simulation cell size in the dimension along the laser axis of propagation.

The results of the simulations are seen to be in very good agreement with the experimental data. As well as correctly reproducing the timescale of the wakefield decay, the wakefield amplitudes calculated by the simulations are found to be close in \emph{absolute} terms to the measured values. For both hydrogen and deuterium plasmas the relative wake amplitude at $\zeta = 0$ was calculated to be $7.5$\% for the laser and plasma conditions of the experiment. This compares with the measured values of $(6 \pm 2)$\% for hydrogen and $(4\pm2)$\% for deuterium. This agreement, which is within a factor of two, is remarkably good when it is considered that: (i) the simulations contained no free parameters; and (ii) the measured wakefield amplitude is rather sensitive to variations in many of the experimental parameters. These experimental parameters include: the pulse energy, duration, and spatio-temporal quality of the laser pulse; the spectrum of the probe pulse; the relative alignment of the drive and diagnostic pulses; and the pressure in the gas cell.

Table II summarises the results of the measurements and simulations; to enable a comparison with the experiments, the temporal variation of the wake amplitude found in the simulations was fitted to an exponential decay. For the conditions of the experiment we find: for hydrogen, $\tau_\mathrm{wf} \approx \SI{9}{ps}$, corresponding to around 76 plasma periods; for deuterium these values are approximately \SI{16}{ps} and 134 periods respectively.

The measured wakefield lifetimes are long compared to the electron plasma period $T_\mathrm{pe} = \SI{121}{fs}$, and are comparable to the ion plasma periods $T_{pi} = \SI{5.2}{ps}$ and \SI{7.4}{ps}  for hydrogen and deuterium respectively. The ratios of the decay times measured for deuterium and hydrogen are found to be $\tau^{D2}_{wf}/\tau^{H_2}_{wf} = 1.78 \pm 0.97$ and $1.76 \pm 0.22$ from the measurements and simulations respectively. The experimentally measured decay times are seen to be in very close agreement with those determined from PIC simulations. Unfortunately the experimental errors, particularly those for $D_2$, are too large to conclude that the ratio $\tau^{D2}_{wf} / \tau^{H2}_{wf}$ differs from unity. However, the PIC simulations do show that the decay time for deuterium is longer than that for hydrogen, and that their ratio is close to, but larger than, the ratio of the ratio of the inverse ion plasma frequencies $\omega_{D2}^{-1} / \omega_{H2}^{-1} = \sqrt{M_{D2}/M_{H2}} = \sqrt{2}$. The measurements and simulations demonstrate that the wake decay time is of the order of the inverse ion plasma period, but that it is too simplistic to assume that the decay time is strictly proportional to this quantity \cite{Boetticher2021}.

A detailed analysis of the PIC simulations \cite{Boetticher2021}, and comparison with work by Sanmartin et al \cite{Sanmartin_1970}, shows plasma waves in these experiments decay via the modulational instability. This instability causes small spatial variations in the ion density to grow exponentially, with a time-scale of order $T_{pi}$, leading to a loss of coherence of the electron oscillations, and hence decay of the wakefield amplitude.

\begin{center}
\begin{table}[tb]
 \caption{Comparison of the wakefield decay times (in picoseconds) obtained from experiments and PIC simulations.}
 \label{tab:decay_time}
 \begin{tabular}{l c c c r} \toprule
 & $\tau^\mathrm{H_2}_\mathrm{wf}$ & $\tau^\mathrm{D_2}_\mathrm{wf}$ &  $\tau^{D_2}_{wf}$/$\tau^{H_2}_{wf}$\\  [0.5ex]
 \hline
\\  [-1.5ex] 
Experiment & $9\pm2$  & $16\pm8$   & $1.78 \pm 0.97$ \\ 
Simulation & $8.5\pm0.9$ &  $15\pm1$ & $1.76\pm0.22$ \\ 
\toprule
\end{tabular}
\end{table}
\end{center}

\section{Conclusion}
In conclusion, we have used single-shot frequency domain holography to measure the lifetime of 1D linear plasma wakefields in hydrogen and deuterium plasmas driven in the short-pulse LWFA regime. Wakefields with relative amplitudes of approximately $6\%$ were driven by \SI{1.7}{J}, \SI{50}{fs} laser pulses in hydrogen and deuterium plasmas of density $n_{e0} = \SI{8.4E17}{cm^{-3}}$. The wakefield lifetimes were measured to be $\tau^\mathrm{H_2}_\mathrm{wf} = (9\pm2)$ ps and $\tau^\mathrm{D_2}_\mathrm{wf} = (16\pm8)$ ps respectively for hydrogen and deuterium. The experimental results were found to be in very good agreement with 2D particle-in-cell simulations.

These findings are of relevance to the MP-LWFA scheme, in which the wakefield is driven resonantly  by a train of short pulses \cite{Hooker_2014, Cowley2017a}. This latter approach is of considerable interest since it offers a route to driving LWFAs at high pulse repetition rates with novel laser technologies which can provide the required \emph{average} power, with high wall-plug  efficiency, but which deliver pulses which are too long to drive a plasma wave directly. The wakefield lifetime is of key importance to the MP-LWFA scheme since it determines the maximum useful number of pulses in the pulse train. The work presented here shows that, for  plasma densities relevant to MP-LWFAs, the wakefield lifetime corresponds to of order 100 plasma periods, which is large compared to the $N \approx 10$ pulses required for MP-LWFA schemes driven by pulse trains of total duration in the picosecond range \cite{Jakobbson:2021}.

\section{Acknowledgements}
This work was supported by the UK Science and Technology Facilities Council (STFC UK) [grant numbers ST/N504233/1, ST/P002048/1, ST/R505006/1, ST/S505833/1, ST/V001655/1]; the Engineering and Physical Sciences Research Council [EP/N509711/1, EP/V006797/1], and the Central Laser Facility. This material is based upon work supported by the Air Force Office of Scientific Research under award number FA9550-18-1-7005. This work was supported by the European Union's Horizon 2020 research and innovation programme under grant agreements No. 653782 and 730871.

This research was funded in whole, or in part, by EPSRC and STFC, which are Plan S funders. For the purpose of Open Access, the author has applied a CC BY public copyright licence to any Author Accepted Manuscript version arising from this submission.

The data associated with this paper, and the  input decks used for the PIC simulations are  available at \url{https://zenodo.org/deposit/7945414}

\bibliography{ion_motion}

\pagebreak

\end{document}


\maketitle
\section{Probe beamline}
\subsection{Experimental setup}
\begin{figure}[h!] 
   \centering
    \includegraphics[width=0.6\columnwidth]{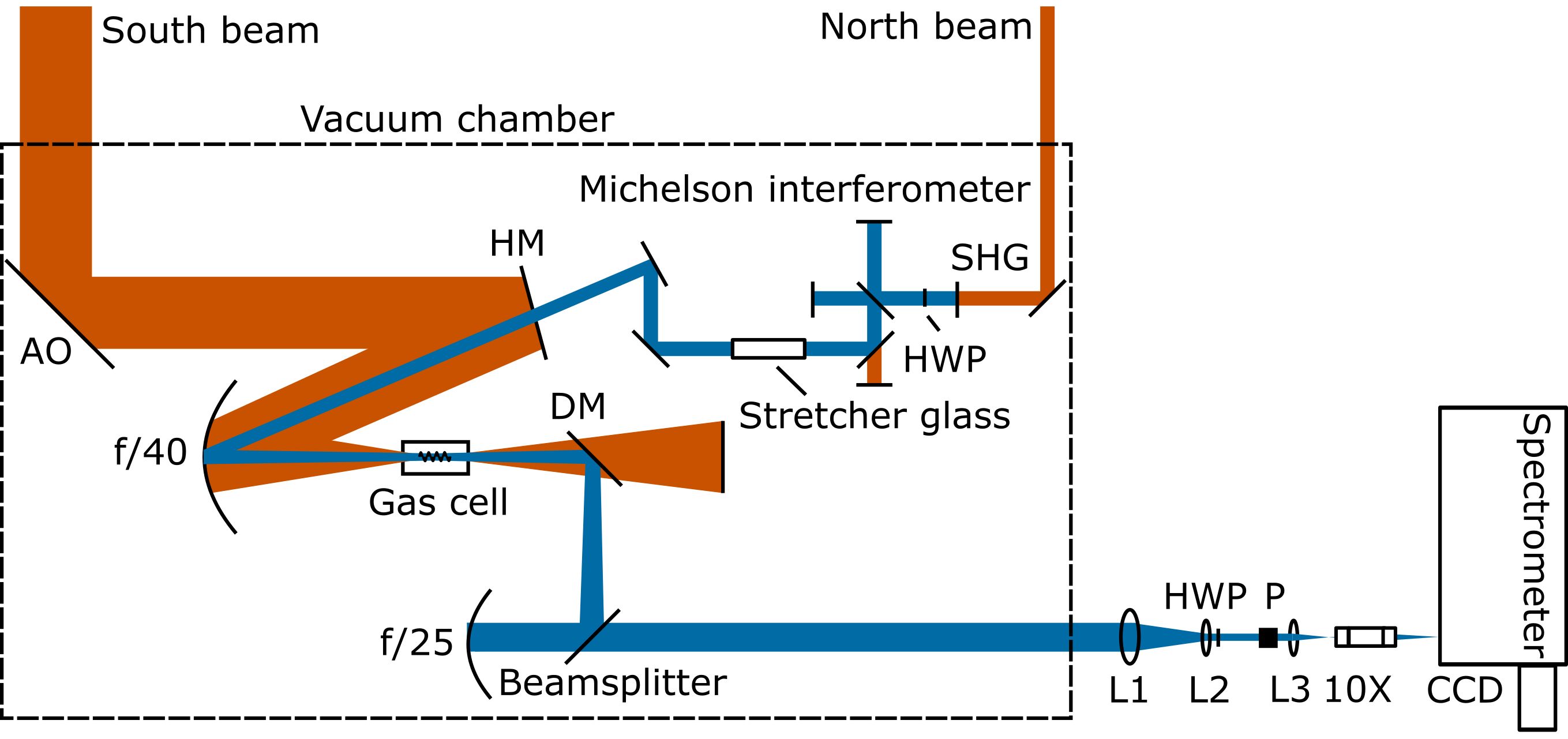}
    \caption{Schematic drawing of the experimental layout inside the target vacuum chamber. Both beams of the  Astra-Gemini TA3 laser were used, one beam used as the drive beam and the other beam as the diagnostic beam. The \SI{800}{\nano\metre} beams are shown in red and the \SI{400}{\nano\metre} diagnostic beam is shown in blue. The diagnostic beam was transported to a \SI{400}{\nano\metre} imaging spectrometer located outside the vacuum chamber. AO: adaptive optic. SHG: second harmonic generating crystal. HWP: half wave plate, P: polarizer, L1: $f=\SI{500}{\milli\m}$ ($4^"$ diameter), L2: $f=\SI{-100}{\milli\m}$ ($1^"$ diameter), L3: $f=\SI{300}{\milli\m}$ ($1^"$ diameter)}
    \label{fig:layout}
\end{figure}   
The diagnostic pulses were generated using one of the Gemini beams (central wavelength $800\mathrm{nm}$), temporally synchronised with the drive beam. This beam was frequency doubled to $\SI{400}{\nano\m}$ and split into two temporally offset pulses using a Michelson interferometer with a path difference corresponding to $\Delta \zeta\approx\SI{6}{\pico\second}$ between the two arms. The pulses were subsequently chirped, and stretched to a pulse duration of approximately \SI{1.35}{\pico\second}, by propagating them through a \SI{160}{\milli\m} long piece of BK7 glass. The diagnostic pulses were injected co-axially with the drive beam by directing them through a holed mirror, and on leaving the gas cell they were reflected by a dichroic mirror and a beamsplitter and directed at normal incidence to a spherical mirror of focal length $f = \SI{2.5}{m}$ ($f/25$). The collimated light from the object plane, located a distance $f$ from the spherical mirror, was returned through an optical wedge onto a set of lenses: L1: $f=\SI{500}{\milli\m}$ and L2: $f=\SI{-100}{\milli\m}$ forming a down-collimating telescope, L3: $f=\SI{300}{\milli\m})$ and a $10X$ microscopic lens ($f=\SI{20}{\milli\m}$) used to image the beam onto the spectrometer. 

A half wave plate (HWP) and polariser setup was used to filter out blue light generated in the plasma by the drive beam.  A simplified drawing of the layout is shown in Figure~\ref{fig:layout}.

The peak intensity of the drive pulse at focus was estimated by using a camera to record the transverse fluence profile of the focus. This was converted to a transverse intensity profile by using the measured energy and duration of the pulse.

\subsection{Spectral phase}\label{eq:measuring_gdd}
\begin{figure}[tb]
\begin{center}
    \includegraphics[width=0.6\columnwidth]{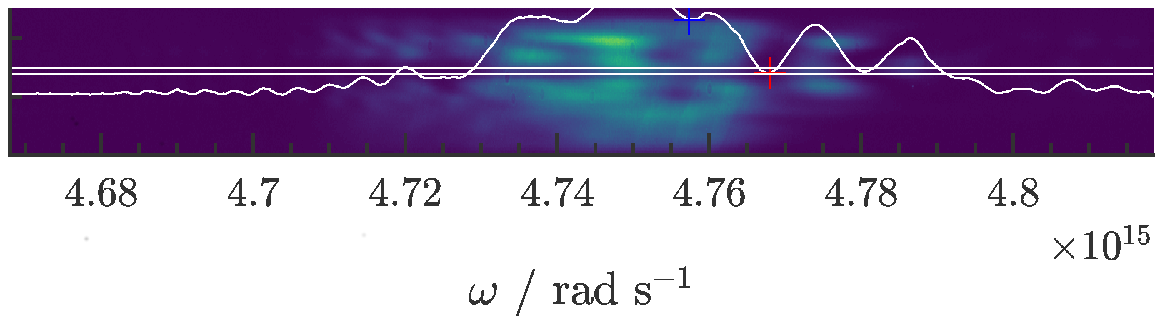}
    \caption{Plot of the perturbed spectrum of the probe pulse, taken during measurements of the spectral chirp of the probe pulse.. The temporal overlap of the probe and drive pulses in the gas cell causes a reduction of the probe pulse intensity as measured on the spectrometer. The red cross marks a low-intensity feature that is tracked as a function of the temporal delay between the drive and probe (or reference) pulses, in order to measure the spectral phase.}
    \label{fig:ionisation_front}
    \end{center}
\end{figure}
The spectral phases of the diagnostic pulses (i.e. the probe and reference pulses) were measured by recording the spectrum of the relevant diagnostic pulse after it had co-propagated through the gas cell with the driver pulse. A sharp reduction in the spectral intensity of the diagnostic pulse was observed at the local frequency corresponding to the point in the probe pulse which overlapped with the drive pulse. This reduction of intensity was caused by the high refractive gradient at the ionization front produced by the drive pulse. By tracking, as a function of the relative delay $\zeta_i$ of the drive pulse, the  frequency at which this intensity reduction occurred,  it was possible to measured the ``local'' frequency $\omega(\zeta_i)$ of the diagnostic pulse. An example measurement is shown in Figure~\ref{fig:ionisation_front}, for one temporal delay. 

The measured frequency $\omega(\zeta_i)$ was used to estimate the spectral phase in the following way. By measuring the probe spectrum separately on a spectrometer we obtained its spectral amplitude $|\mathcal{E}(\omega)|$, from which we could reconstruct the spectral representation of its electric field as $\mathcal{E}(\omega)=|\mathcal{E}(\omega)|\exp[\varphi(\omega)]$. Here $\varphi(\omega)$ is the spectral phase to be estimated, assumed to be a third-order polynomial: 
$$\varphi(\omega)=\varphi_0+\varphi_1(\omega-\omega_0)+\frac{1}{2}\varphi_2(\omega-\omega_0)^2+\frac{1}{6}\varphi_3(\omega-\omega_0)^3,$$
where the coefficients $\varphi_1$, $\varphi_2$, and $\varphi_3$ are to be determined. By guessing initial values for these coefficients, the temporal representation of the electric field can be obtained from the Fourier transform of $\mathcal{E}(\omega)$, $E(\zeta)=\operatorname{FT}[ \mathcal{E}(\omega)]$. The temporal phase is obtained from its argument: $\phi(\zeta)=\mathrm{arg}[E(\zeta)]$. Noting that the instantaneous frequency is given by the derivative of the temporal phase, $\omega(\zeta)=d\phi(\zeta)/d\zeta$, we compared this reconstructed frequency with the measured frequency $\omega(\zeta_i)$ using the overlap technique described above. We could improve on the guess of the the spectral phase $\varphi(\omega)$ and find the value that best matched the measured instantaneous frequency by minimising the squared difference between these:
\begin{align}
  \min{\sum_i\left(\left|\omega_\mathrm{probe, ref}(\zeta_i) - \frac{d\phi(\zeta_i)}{d\zeta}\right|^2\right)}.
  \label{eq:varphi_min}
\end{align}

\begin{figure}[h!]
\centering
\begin{subfigure}{.5\textwidth}
\centering
\includegraphics[width=6.5cm]{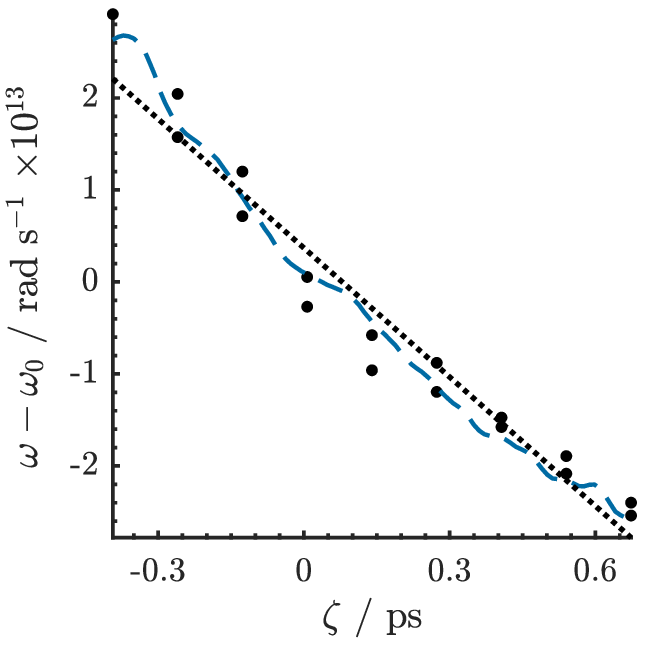}
\caption{}
\label{fig:gdd_probe}
\end{subfigure}%
\begin{subfigure}{.5\textwidth}
  \centering
\includegraphics[width=6.5cm]{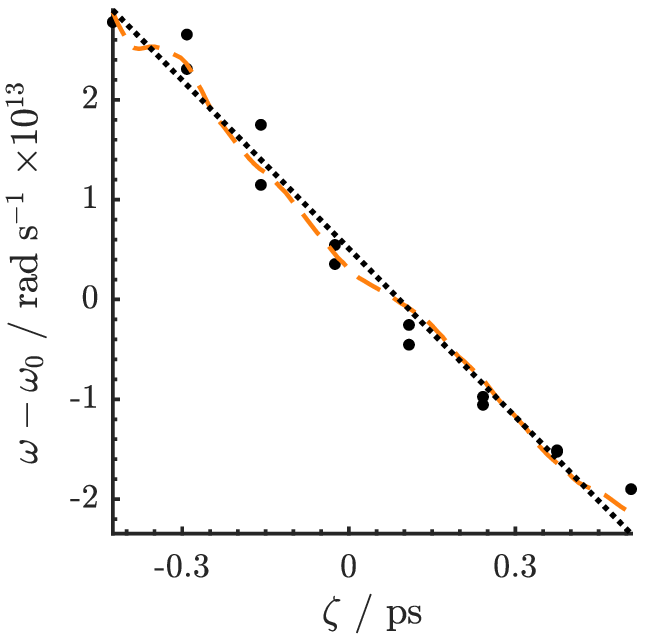}
\caption{}
\label{fig:gdd_ref}
\end{subfigure}
\caption[GDD measurement of the probe pulses]{Measurement of the spectral phase of the (a) probe and (b) reference pulses. The black dots indicate the location of the ionisation front feature in time and frequency. The dashed lines show the fitted first derivative of the temporal phase $d\phi(\zeta)/d\zeta$ of each pulse. The dotted lines show a linear fit to the data.}
\label{fig:gdd_measurement}
\end{figure}

Using this technique, $\varphi_2$ (i.e. the GDD) was found to be \SI{18600\pm790}{\femto\s\squared} and \SI{17600\pm800}{\femto\s\squared} for the probe (Figure \ref{fig:gdd_probe}) and references pulses (Figure \ref{fig:gdd_ref}), respectively. These values were used in the TESS analysis.
\section{The TESS analysis method}
\begin{figure}
\centering
\begin{subfigure}{.5\textwidth}
  \centering
 \includegraphics[width=6.5cm]{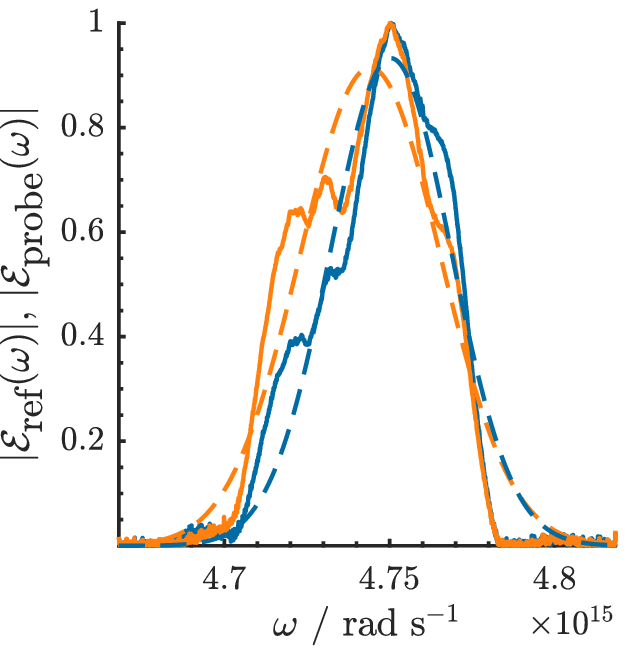}
\caption{}
\label{fig:spec}
\end{subfigure}%
\begin{subfigure}{.5\textwidth}
  \centering
\includegraphics[width=6.5cm]{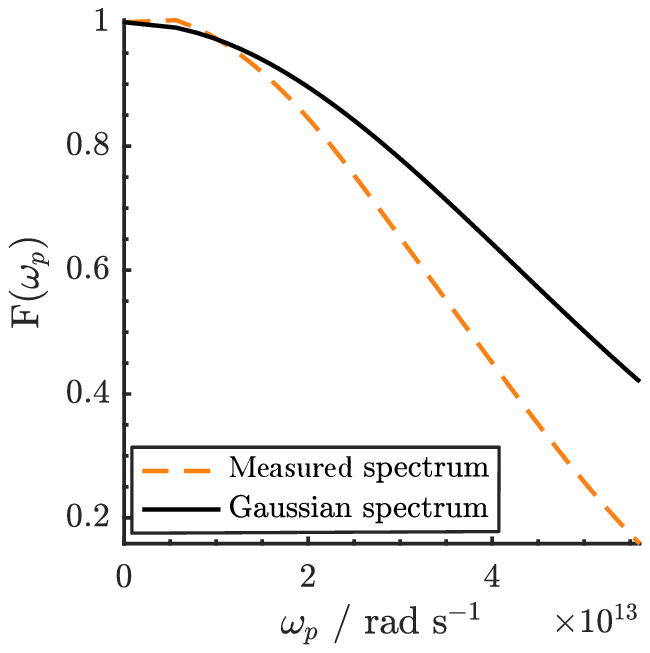}
\caption{}
\label{fig:overlap}
\end{subfigure}
\caption[Spectral overlap]{(a) Lineouts of the spectral intensities of the probe (red line) and reference pulses (blue line). The dashed lines show Gaussian fits to these spectra. (b) spectral overlap function $\mathcal{F}(\omega_{pe})$ as a function of the electron density, for the measured spectra and Gaussian spectra fitted to the measured spectra.}
\label{fig:spec_overlap}
\end{figure}

TESS is a simplified analysis method to obtain the wakefield amplitude from spectral interferometric data obtained using an FDH diagnostic setup. Following Matlis et al. \cite{Matlis2016a}, let a propagating reference pulse, ahead in time of the pump and probe pulses, and a probe pulse co-propagating with the wakefield be described by $E_{\text{ref}}(\zeta)$ and the phase-modulated probe pulse by $ E^\prime_{\text{probe}}(\zeta) = E_{\text{probe}}(\zeta)\exp[i\phi_0\sin(\omega_{pe}\zeta)]$, where the wakefield is assumed to be sinusoidal with $\phi_0$ being the phase amplitude of the sinusoidal wave, and $\omega_{pe}$ the plasma frequency. Taking the Fourier transform of the probe pulse one obtains 
\begin{align}
\mathcal{E}^\prime_{\text{probe}}(\omega) &= \frac{1}{\sqrt{2\pi}}\int_{-\infty}^{\infty}E_{\text{probe}}(\zeta)\exp[i\phi_0\sin(\omega_{pe}\zeta)]\exp(-i\omega \zeta)d \zeta.\\
  &= \frac{1}{\sqrt{2\pi}}\int_{-\infty}^{\infty}E_{\text{probe}}(\zeta)\sum_{-\infty}^{\infty}J_k(\phi_0)\exp(ik\omega_{pe}\zeta)\exp(-i\omega \zeta)d \zeta\\ 
  &=\sum_{k=-\infty}^{\infty}J_k(\phi_0)\mathcal{E}_{\text{probe}}(\omega-k\omega_{pe}),
  \label{eq:FTprobe1}
\end{align}
using the Jacobi-Anger expansion to express the phase term as an infinite sum of Bessel functions of the first kind:
\begin{equation}
\exp[i\phi_0\sin(\omega_{pe}\zeta)] = \sum_{k=-\infty}^{\infty}J_k(\phi_0)\exp(ik\omega_{pe}\zeta).
\end{equation}
The spectral interference intensity between the probe and reference pulses is given by: 
\begin{equation}
S(\omega) = |\mathcal{E}_{\text{probe}}(\omega)+\mathcal{E}_{\text{ref}}(\omega)|^2.
\end{equation}
Using Equation \eqref{eq:FTprobe1}, this can be expanded into a sum of the cross-terms of the probe and reference spectra, as well as the different copies contained in the probe spectrum:
\begin{align}
  S(\omega) &=|\mathcal{E}_{\text{probe}}(\omega)|^2+|\mathcal{E}_{\text{ref}}(\omega)|^2+\mathcal{E}^{*}_{\text{probe}}(\omega)\mathcal{E}_{\text{ref}}(\omega)+\text{c.c.}\\
&=\sum_{m,n}J_n(\phi_0)J_m(\phi_0)\mathcal{E}^{*}_{\text{probe}}(\omega-n\omega_{pe})\mathcal{E}_{\text{probe}}(\omega-m\omega_{pe}) + |\mathcal{E}_{\text{ref}}(\omega)|^2\label{eq:Sw1} \\
&+ \sum_kJ_k(\phi_0)\mathcal{E}^{*}_{\text{ref}}(\omega)\mathcal{E}_{\text{probe}}(\omega-k\omega_{pe})\exp(-i\omega\Delta \zeta) + \text{c.c.}\nonumber,
\end{align}
where c.c. stands for the complex conjugate of the previous term, and the Fourier shift theorem was used to introduce $\Delta\zeta$, the temporal delay between the probe and reference pulses. The Fourier transform of the spectral interferogram in Eq.~\eqref{eq:Sw1} is given by:
\begin{equation}
S(\zeta) = \sum_mg_m(\phi_0,\zeta,\omega_{pe})H(\zeta,m\omega_{pe}) + H(\zeta,0) + \sum_kJ_k(\phi_0)H(\zeta-\Delta \zeta,k\omega_{pe}) + \text{c.c.},\label{eq:St1}
\end{equation}
where the function $H$ is given by:
\begin{equation}
H(\zeta,\omega_{pe}) = \frac{1}{\sqrt{2\pi}}\int_{-\infty}^{\infty}\mathcal{E}^*_{\text{ref}}(\omega)\mathcal{E}_{\text{probe}}(\omega-\omega_{pe})\exp(i\omega \zeta) d\omega,
\end{equation}
and the function $g$ by:
\begin{equation}
g_m(\phi_0, \zeta, \omega_{pe}) = \sum_nJ_n(\phi_0)J_{m+n}(\phi_0)\exp(in\omega_{pe}\zeta).
\end{equation}
This can be rewritten as, defining $H_0(\zeta) \equiv H(\zeta,\omega_{pe}=0)$
\begin{align}
  S(\zeta) &= \underbrace{\sum_mg_m(\phi_0,\zeta,\omega_{pe})f(m\omega_{pe})H_0(\zeta+m\tau) + H_0(\zeta)}_{\text{DC peak}}\\
  &+ \underbrace{\sum_kJ_k(\phi_0)\mathcal{F}(k\omega_{pe})H_0(t-\Delta \zeta+k\tau) + \text{c.c.}}_{\text{Sidebands+satellites}},
  \label{eq:sideband}
\end{align}
where $\tau\equiv\varphi^{(2)}\omega_{pe}$, and $\mathcal{F}(k\omega_{pe})$ is a spectral overlap factor, which in general is given by \cite{Arran2018a}:
\begin{align}
  \mathcal{F}(k\omega_{pe}) = \frac{\int_{-\infty}^{\infty}|\mathcal{E}_{\text{probe}}(\omega + k\omega_{p})||\mathcal{E}_{\text{ref}}(\omega)|d\omega}{\int_{-\infty}^{\infty}|\mathcal{E}_{\text{probe}}(\omega)||\mathcal{E}_{\text{ref}}(\omega)|d\omega}.
  \label{eq:spectral_overlap}
\end{align}
The measured probe and reference spectra and their spectral overlap factors are plotted in Figure~\ref{fig:spec_overlap}. 

\begin{figure}[t!]
\begin{center}
    \includegraphics[width=0.5\columnwidth]{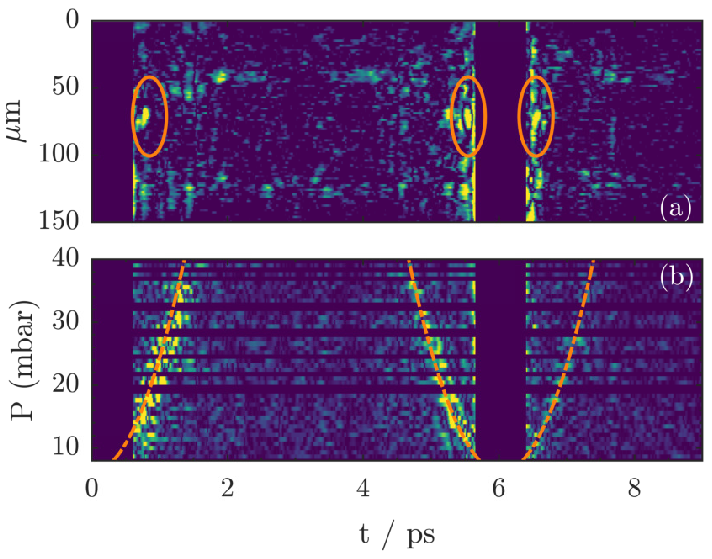}
    \caption{(Top) Intensity of the Fourier transform of the recorded spectral interferograms. The three TESS satellites on either side of the sideband at $t = \Delta \zeta = 6$ ps, and next to the DC band at $t = 0$ ps are indicated. (Bottom) Waterfall plot of Fourier transformed interferograms for a range of cell pressures $P$. For each Fourier transform a \SI{1.7}{\micro\m} wide strip, centred on the peak of the satellite, is shown. The dashed line shows a parabolic fit to the satellite-to-sideband separation, using the satellites on the left side of the sideband (see §3). The above plots are shown using a logarithmic intensity scale.}
    \label{fig:TESS}
    \end{center}
\end{figure}

The ``sideband'' term in equation \eqref{eq:sideband} contains TESS satellite intensity peaks ($k\neq0$) and the sideband caused by interference ($k=0$). These peaks are plotted in Figure \ref{fig:TESS}. In this figure is also shown the dependence on the temporal separation between the TESS satellite peaks and the sideband term with plasma density, due to variation of the $\omega_{pe}$ term in $\tau$. The ratio $r$ between the first TESS satellite in equation \eqref{eq:sideband} ($k=1$ term, evaluated at $\zeta=\Delta \zeta - \tau$) and the sideband ($k=0$ term, evaluated at $\zeta=\Delta \zeta$), is:
\begin{align}
  r = \frac{J_1(\phi_0)\mathcal{F}(\omega_{pe})}{J_0(\phi_0)}.
\end{align}
For small amplitudes $\phi_0\ll1$, the Bessel functions of the first kind $J_{\alpha}$ can be expanded as:
\begin{align}
  J_\alpha(\phi_0) &\approx \frac{1}{(\alpha-1)!}\left(\frac{\phi_0}{2}\right)^\alpha\label{eq:Bessel_exp},
\end{align}
with  $J_0(\phi_0) \approx 1$ and $J_1(\phi_0) \approx \phi_0/2$. The phase amplitude $\phi_0$ can therefore be expressed as:
\begin{align}
  \phi_0 = \frac{2r}{\mathcal{F}(\omega_{pe})}.
  \label{eq:phi0_tess}
\end{align}

The plasma wave is assumed to be sinusoidal:
\begin{align}
  n_e(\zeta) = n_{e0} + \delta n_e\sin(\omega_{pe}\zeta).
  \label{eq:ne1}
\end{align}
The phase difference acquired by a laser of wavelength $\lambda$ propagating through plasma is given by:
\begin{align}
  \Delta\phi(\zeta) = \int\frac{n_e(\zeta)e^2\lambda}{4\pi\epsilon_0m_e}dz.
  \label{eq:deltaphi1}
\end{align}
Inserting Equation \eqref{eq:ne1} and removing the constant density term yields:
\begin{align}
  \Delta\phi(\zeta) = \int\frac{\omega_{pe}^2\lambda}{4\pi}\frac{\delta n_e}{n_{e0}}\sin(\omega_{pe}\zeta) d z,
  \label{eq:deltaphi3}
\end{align}
with $\omega_{pe}^2/n_{e0} = e^2/\epsilon_0m_e$. Assuming that there is no significant longitudinal variation this can be integrated over the length $L$ of the plasma:
\begin{align}
  \Delta\phi(\zeta) = \frac{\omega_{pe}^2\lambda}{4\pi}\frac{\delta n_e}{n_{e0}}\sin(\omega_{pe}\zeta)L.
  \label{eq:deltaphi4}
\end{align}
The phase amplitude is thus identified as:
\begin{align}
\phi_0 = \frac{\omega_{pe}^2\lambda}{4\pi}\frac{\delta n_e}{n_{e0}}L.
\end{align}
Using Equation \eqref{eq:phi0_tess} the relative wakefield amplitude $\delta n_e/n_{e0}$ is given by:
\begin{align}
  \frac{\delta n_e}{n_{e0}} &=\phi_0\frac{4\pi}{\omega_{pe}^2\lambda L} \\
  &=\frac{2r}{\mathcal{F}(\omega_{pe})}\frac{4\pi}{\omega_{pe}^2\lambda L}.
  \label{eq:tess_wa}
\end{align}
\section{Pressure scan}

\begin{figure}[h!]
\begin{centering}
\includegraphics[width=8cm]{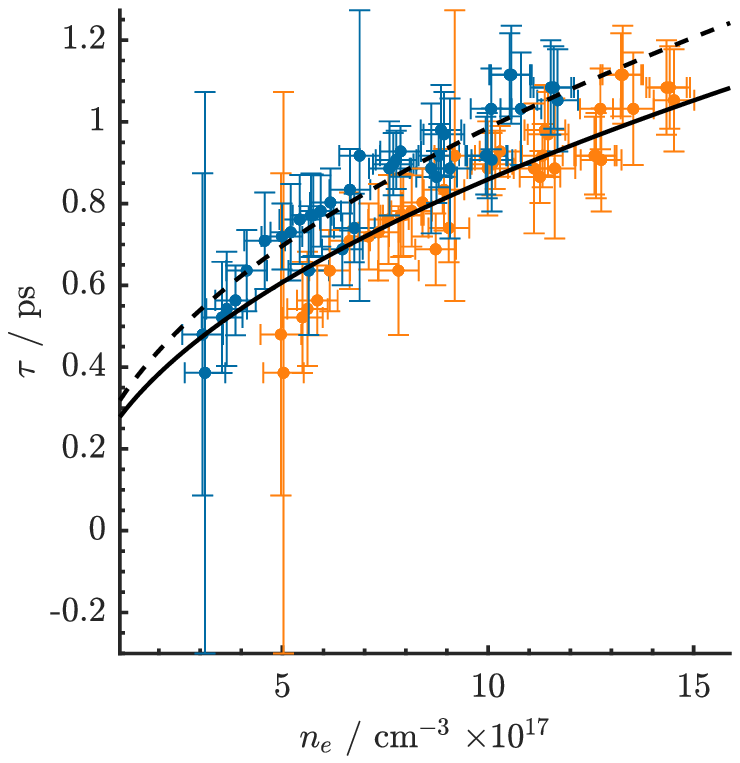}
\caption[Peak separation plot]{Plot showing a fit to the temporal separation $\tau$ between the sideband and the TESS satellites as a function of plasma density. The red points assume that the pressure in the cell is that measured by the gauge, $P_\text{meas}$; the blue points assume that the true pressure in the cell is given by $P_\text{cell} = \alpha P_\text{meas} - P_0$. For both plots that the electron density is calculated assuming complete ionization of the H2 gas at the corresponding pressure. The horizontal error bars correspond to the pressure resolution of $1$ mbar of the pressure gauge, and the vertical bars show the estimated error of the TESS satellite location. The solid (dashed) line shows a fit to the red (blue) points of the form $\tau=\omega_{pe}\varphi^{(2)}$.}
\label{fig:peak_sep}
\end{centering}
\end{figure}

In addition to varying the temporal delay $\zeta$ to measure the wakefield amplitude over time, in a separate experiment the pressure of hydrogen was was varied between $10$ mbar and $40$ mbar. A one-pixel wide slice in the $y$-direction centred on the TESS peak was selected for each shot, and used to populate the waterfall plot in Figure~\ref{fig:TESS}. The positions of the peaks relative to the sideband were then measured as a function of plasma density. This data was used to fit the expression $\tau = \varphi^{(2)}\omega_{pe}$, using the GDD ($\varphi^{(2)}$) as the fitting parameter. Fitting directly to the data, a GDD value of $\varphi^{(2)}=\SI{15203\pm327}{\femto\s\squared}$ was obtained. However, it was observed that the data appeared to be offset from this simple fit, as can be seen in Figure~\ref{fig:peak_sep}. This is likely caused by a drop in pressure between the location of the pressure gauge, which was connected to the gas pipe at a distance of $\sim\SI{2}{\m}$ from the cell. A second fit was therefore performed, where the pressure in the cell was assumed to be given by: $P_\text{cell} = \alpha (P_\text{gauge} - P_0)$, where the offset $P_0$ accounts for any potential calibration offset.With this method the fitted parameters were found to be $\alpha=0.96$, $P_0=\SI{3}{\milli\bar}$, and $\varphi^{(2)}=\SI{16946\pm81}{\femto\s\squared}$, which is also shown in Figure~\ref{fig:peak_sep} (where the offset and shift parameters have been applied to the measured data rather than to the fit). This value of the GDD is within $\sim10 \%$ of that obtained using the spectral blowout technique described in Section~\ref{eq:measuring_gdd}, confirming the that the procedure adopted for relating the cell pressure to the gauge pressure is correct.

\section{Particle-in-cell simulations}\label{sec:PIC}
Numerical simulations of the experiments were performed with the relativistic particle-in-cell code \texttt{smilei} \cite{SMILEI} on the ARCHER UK National Supercomputing Service and the Oxford Advanced Research Computing (ARC) Arcus-B cluster, using the same parameters as used in the experiment (the parameters used in the simulations are summarized in Table \ref{table:sim}). In the simulations, the plasma has a longitudinal and transverse extent of 240 $\mu$m ($8\lambda_p$) and 160 $\mu$m, respectively, at a uniform plasma density, $n_{e0} = 9.7\times 10^{17}$ cm$^{-3}$. The simulation window was fixed, and boundaries were absorptive to electromagnetic radiation and reflective to electron and ion particles. The wakefield amplitude was calculated from that of the electric field of the Fourier mode corresponding to the wakefield plasma oscillation. Modulations of the longitudinal electric field developed within two picoseconds after excitation in the simulations, on a scale much smaller than the wakefield wavelength ($\sim 1.3$ $\mu$m versus $\lambda_p \approx 30$ $\mu$m). 

\begin{table}[!htb]
\centering
\caption{Parameters of plasma, and drive and probe laser beams.}
\scriptsize
\begin{tabular}{lr}
	Simulation laser and plasma parameters &  \\
	\midrule
	\midrule
	\textit{Drive pulse parameters:} & \\ 
	Wavelength [nm] & 800 \\
	Pulse duration [FWHM] [fs] & 40 \\
	Pulse width [FWHM] [$\mu$m] & 45 \\
	Peak intensity [Wm$^{-2}$] & $1\times10^{18}$  \\
	$a_0$ & 0.5  \\
	\textit{Plasma parameters:} & \\ 
 	$n_e$ [$m^{-3}$]& \SI{9.7e23}{}   \\
 	$\lambda_D$ [$\mu m$]  & 0.025   \\
	$\Delta x$ [$\lambda_D$]  & 1    \\
	$\Delta y$ [$\lambda_D$]  & 5  \\
	$\Delta t$ [$t_{\mathrm{CFL}}]$& 0.95  \\
	$n_\mathrm{ppc}$    & 64  \\
	$x_{\mathrm{max}}$ [$\lambda_p$]& 8     \\
	$y_{\mathrm{max}}$ [$\mu m$] & 160 \\
	\midrule
	\midrule
\end{tabular}

\label{table:sim}
\end{table}

%
\section{Estimating the plasma temperature}
\begin{table}[!htp]\centering
\caption{Simulation parameters used in 2D PIC simulations to estimate the plasma electron temperature.}\label{tab:PICTemp}
\scriptsize
\begin{tabular}{llr}
Parameter &Value &Unit \\\midrule
\midrule
Particles per cell &$32$ &- \\
Simulation length &$150$ &$\mu$m \\
Simulation width &$300$ &$\mu$m \\
Resolution (along laser axis) &$0.05$ &$\lambda$ \\
Resolution (perpendicular to laser axis) &$0.1$ &$\lambda$ \\
Gas species &H$_2$ &- \\
Plasma density &$\SI{9.7e17}{}$ &cm$^{-3}$ \\
Laser temporal duration &$\SI{48.9}{}$ &fs \\
Laser spot width &$\SI{52.3}{}$ &$\mu$m \\
Laser intensity &$\SI{6.5e17}{}$ &Wcm$^{-2}$ \\
\midrule
\midrule
\end{tabular}
\end{table}
In order to estimate the parameter $W$, which determines whether the modulational instability is in the strong or weak field regime, for the experimental conditions described in this paper we need an estimate of the temperature after ionisation in the plasma. To obtain this estimate, we performed 2D particle-in-cell (PIC) simulations in the code Extendable PIC Open Collaboration (EPOCH) \cite{Arber2015}. EPOCH includes ionisation models, which take into account both tunnelling and barrier suppression ionisation pathways \cite{Douglas2013}. Collisional ionisation is a third pathway that is less important in the case of short-pulse laser ionisation; with a measured ionisation cross-section of hydrogen of $\sigma\approx \SI{5e-17}{\square\cm}$ \cite{Shah1987} for non-relativistic electrons with quiver velocities $v_{\text{osc}}=a_0c \approx 0.5c$ (with $a_0\approx 0.5$), the ionisation rate is approximately given by:
\begin{align}
W = n_e\sigma v_{\text{osc}} [\SI{}{\per\s}] \approx \SI{0.72}{\per\pico\s}.
\end{align}
Since collisional ionisation is important over the scale of picoseconds rather than femtoseconds, as is the case for short-pulse laser ionisation, we neglected this contribution in the simulations. The simulation parameters, which were chosen to be equal to the experimental parameters, are summarised in Table \ref{tab:PICTemp}. Two different methods were used for estimating the temperature from these simulations. In the first, which is a built-in temperature probe in EPOCH, the temperature is approximated as the standard deviation of the total momentum in each simulation cell $i$,
\begin{align}
  k_bT_i \approx \frac{\langle p^2\rangle_i}{2m}.
\end{align}
 Averaging along the axis of laser propagation, we obtained an average on-axis temperature of $2.75$ eV. The second way of approximating the temperature is by measuring the random (thermal) motion of the electrons, from which we obtained the thermal energy distribution $E_\text{thermal} = p_{\text{thermal}}^2/2m$. During the time that the drive laser spends in the plasma slab, there is a strong coherent quiver motion at the laser frequency. By filtering out this low-frequency component from the momentum we are left with only the thermal component from which we estimated the thermal energy distribution. The temperature was obtained using a fit of the Maxwell-Boltzmann distribution, with the temperature $T$ as the fitting parameter:
\begin{align}
f_{E}(E, T)=2\sqrt{\frac{E}{\pi}}\left(\frac{1}{k_B T}\right)^{3 / 2} \exp \left(\frac{-E}{k_B T}\right).
\end{align} 
The laser-ionisation processes investigated in this paper do not initially lead to a thermal distribution. Over time however, the electrons will equilibrate to a thermal distribution through collisions. We found that the simulation results could be well characterised as a sum of two approximately equal electron populations with different temperatures $T_1$ and $T_2$, with $T_1=0.26$ eV and $T_2=1.54$ eV. Coulomb collisions between these two populations lead to a thermal distribution after a short time. This equilibration process may be described by \cite{Melrose2011}:
\begin{align}
\frac{d T_{\alpha}}{d t}=\sum_{\beta} \bar{\nu}_{\epsilon}^{\alpha \backslash \beta}\left(T_{\beta}-T_{\alpha}\right),
\end{align}
where the Spitzer collision frequency is given by:
\begin{align}
\bar{\nu}_{\epsilon}^{\alpha \backslash \beta}=1.8 \times 10^{-19} \frac{\left(m_{\alpha} m_{\beta}\right)^{1 / 2} Z_{\alpha}^{2} Z_{\beta}^{2} n_{\beta}\ln\lambda}{\left(m_{\alpha} T_{\beta}+m_{\beta} T_{\alpha}\right)^{3 / 2}} \SI{}{\per\s}.
\end{align}  
For electron-electron equilibration $m_{\alpha, \beta} = m_e$, $Z_{\alpha}=Z_{\beta}=1$. Since it was found that the populations were approximately equal in size, we used $n_{\alpha}=n_{\beta}=0.5 n_e$. 

The Coulomb logarithm for these parameters is $\ln\lambda\approx 13.7$. The equilibration time was defined as \cite{Melrose2011}
\begin{align}
\tau_{\epsilon}^{\alpha / \beta} \equiv \frac{1}{\bar{\nu}_{\epsilon}^{\alpha \backslash \beta}} \approx 1.7 \times 10^{5} \frac{\left(T_{\alpha}[\mathrm{eV}]+T_{\beta}[\mathrm{eV}]\right)^{\frac{3}{2}}}{n_{\beta}\left[\mathrm{cm}^{-3}\right] \ln\lambda}\SI{}{\s}\approx\SI{0.1}{\pico\s}.
\end{align}
With the low temperature obtained in the simulation, one sees that $\tau\ll 1$ ps. The two different plasma electron will also become isotropic on a similar timescale \cite{SpitzerJr.2006}, and will therefore be fully thermalised on a time scale much shorter than the picosecond timescale relevant for the wakefield decay process. The final electron temperature used was the average between the two methods outlined in this section, $T = 2.75/2 + (0.26/2 + 1.54/2)/2  \approx 2$ eV. 

\section{Acknowledgements}
This research was funded in whole, or in part, by EPSRC and STFC, which are Plan S funders. For the purpose of Open Access, the author has applied a CC BY public copyright licence to any Author Accepted Manuscript version arising from this submission.

\printbibliography[heading=bibintoc,title=References]